\documentclass[prc,reprint,superscriptaddress,floatfix]{revtex4-2}

\usepackage{amsmath,amssymb,bm,graphicx}
\usepackage[colorlinks=true,linkcolor=blue,citecolor=blue,
 urlcolor=blue]{hyperref}

\newcommand{\HeFour}{\ensuremath{{}^{4}\mathrm{He}}}
\newcommand{\tHp}{\ensuremath{{}^{3}\mathrm{H}+p}}
\newcommand{\tHen}{\ensuremath{{}^{3}\mathrm{He}+n}}

\newcommand{\Smat}{\ensuremath{\mathcal{S}}}
\newcommand{\Kmat}{\ensuremath{\mathcal{K}}}
\newcommand{\fm}{\ensuremath{\mathrm{fm}}}
\newcommand{\MeV}{\ensuremath{\mathrm{MeV}}}

\begin{document}

\title{Coupled-channel scattering from artificial confinement}

\author{Tafat Weiss Attia}
\affiliation{The Racah Institute of Physics, The Hebrew University,
Jerusalem 9190401, Israel}
\author{Itay Horin}
\affiliation{The Racah Institute of Physics, The Hebrew University,
Jerusalem 9190401, Israel}
\author{Betzalel Bazak}
\email{betzalel.bazak@mail.huji.ac.il}
\affiliation{The Racah Institute of Physics, The Hebrew University,
Jerusalem 9190401, Israel}

\date{\today}

\begin{abstract}
Artificial confinement encodes continuum scattering information in discrete,
bound-state-like spectra, allowing reaction observables to be extracted with
finite-basis or finite-domain methods. We apply this strategy to a two-channel
cluster model of $^4$He with open $^3$H+p and $^3$He+n channels. We extract
coupled-channel observables from spectra generated by a harmonic-oscillator
(HO) trap, a spherical hard wall, and, within a single-partial-wave
truncation, a periodic cubic box. The three geometries are formulated in a
unified quantization-condition framework and benchmarked against a continuum
$R$-matrix calculation. Above the second-channel threshold, several confined
levels at a common scattering energy are combined in an overdetermined fit to
determine two phase shifts and an inelasticity. Without Coulomb interactions,
all three geometries yield consistent results for the $^1S_0$ and $^3P_1$
partial waves. With Coulomb interactions in the charged $^3$H+p channel, the
HO and spherical-wall results also agree closely with the continuum
reference. A Monte Carlo propagation study shows that spectral uncertainties
are amplified near trap-function poles and along poorly conditioned
directions associated with the inelasticity and phase-shift difference,
whereas the phase-shift sum remains comparatively robust. These results
provide a controlled benchmark for confinement-based scattering methods and
delineate their strengths and limitations for future few-body and
\textit{ab initio} reaction calculations.
\end{abstract}

\maketitle

%=====================
\section{Introduction}
%=====================

\textit{Ab initio} calculations of nuclear reactions are essential for
connecting observables to the underlying interparticle interactions. They
are, however, substantially more challenging than bound-state calculations.
Scattering observables are defined by continuum boundary conditions and
asymptotic wave functions, whereas many accurate few- and many-body methods
are naturally formulated in finite bases or spatial domains and therefore
produce discrete, bound-state-like spectra.

Artificial confinement provides a bridge from such discrete calculations to
continuum observables. A known external trap or boundary condition discretizes
the continuum spectrum, and a geometry-dependent quantization condition (QC)
relates the resulting energy levels to the infinite-volume scattering matrix.
For two open channels, however, the spectrum-to-scattering inversion is
underdetermined at a fixed energy: the physical scattering matrix contains two
channel phase shifts and an inelasticity, whereas each confined level supplies
only one nonlinear constraint. Recovering all three observables therefore
requires several independent level crossings at the same target energy, an
inverse problem absent from a single-channel extraction.

The most extensively developed finite-volume realization is a periodic cubic
box. L\"uscher showed that two-particle finite-volume energies encode elastic
scattering phase shifts~\cite{Luscher1986,Luscher1991}, and the formalism was
subsequently extended to multiple open channels, arbitrary spin, moving frames,
and higher partial waves
\cite{He2005,Guo2013,Briceno2014,LuuSavage2011,Morningstar2017}. Because a
cubic box preserves only the cubic subgroup of the rotational group, partial
waves subduced into the same cubic irreducible representation can mix.

A second realization imposes a spherical hard wall outside the interaction
region. This geometry preserves angular momentum and leads to a direct radial
matching condition. The spherical-wall method was introduced for nuclear
lattice scattering and later refined to improve the extraction of phase shifts
and mixing angles~\cite{Borasoy2007,Lu2016}. Bovermann \textit{et al.}
generalized the construction to an arbitrary number of coupled channels by
adding an auxiliary channel-mixing interaction, thereby enabling reconstruction
of the full scattering matrix~\cite{Bovermann2019}.

Harmonic-oscillator (HO) confinement is especially attractive for nuclear
calculations formulated in oscillator bases. Busch, Englert, Rza\.{z}ewski,
and Wilkens related the trapped spectrum of two particles with a zero-range
$s$-wave interaction to free-space scattering parameters~\cite{Busch1998}.
Subsequent works extended the approach to higher partial waves and finite-range
interactions, applied it to nucleon--nucleon scattering, and implemented it in
\textit{ab initio} and few-nucleon calculations
\cite{Suzuki2009,Luu2010,BazakEliyahu2016,XilinZhang2020,ZhangPRL2020,SchaferBazak2023,
BagnarolFiveBody2023,AvrahamBazak2026}.

Long-range Coulomb interactions require additional care because the standard
short-range QCs must be modified. Finite-volume and more general
artificial-confinement formulations for charged particles were derived by Guo
and collaborators~\cite{GuoGasparian2021,Guo2021}. For HO confinement, Zhang
\textit{et al.} evaluated the Coulomb contribution through a perturbative
expansion of the relevant Green functions~\cite{ZhangCharged2024}, whereas
Bagnarol \textit{et al.} developed stabilized iterative and direct-inversion
methods for $\ell=0$ and $1$~\cite{Bagnarol2025}. The direct method evaluates
the trap function reliably on either side of the noninteracting HO poles. In a
spherical-wall geometry, the charged-channel asymptotic wave function can
instead be matched directly to regular and irregular Coulomb functions.

The coupled-channel generalization of the periodic finite-volume QC is well
established~\cite{He2005,Guo2013,Briceno2014,Morningstar2017}. For HO
confinement, Guo and Long derived the corresponding relation between the
confined spectrum and the coupled-channel scattering matrix~\cite{GuoLong2022},
and Zhang \textit{et al.} incorporated Coulomb effects and applied the method
to the \tHp--\tHen\ cluster model of \HeFour\ \cite{ZhangCoupled2024}.
Integrated confined correlation functions provide a complementary route and
have also been extended to coupled channels and Coulomb interactions
\cite{GuoGasparian2023,GuoLee2025,GuoLeeCoulomb2025}. Here we instead use the
spectrum-based QC to extract two channel phase shifts and an inelasticity from
an overdetermined set of individual confined levels.

A central question is whether different confinement geometries yield the same
coupled-channel observables when applied to a common Hamiltonian, and how the
inverse extraction responds to Coulomb interactions, trap-function poles,
spectral interpolation, and finite precision in the confined energies.
The \tHp--\tHen\ model provides a useful benchmark because it combines a nearby
second-channel threshold with Coulomb interactions in only one channel. The
model has been studied using complex-scaling and continuum-level-density
methods~\cite{Suzuki2008,Odsuren2021}, and its realization in an HO trap was
examined in Ref.~\cite{ZhangCoupled2024}. We obtain the continuum reference for
the same Hamiltonian using the calculable $R$-matrix method
\cite{LT1958,Descouvemont2010}.

Building on the earlier HO study, we place the HO trap, spherical hard wall,
and periodic cubic box in a common QC framework. For the $^1S_0$ and $^3P_1$
partial waves, we extract the two-channel scattering observables and benchmark
each geometry against the same $R$-matrix calculation. Without Coulomb
interactions, all three geometries yield mutually consistent results and
reproduce the continuum benchmark. With Coulomb interactions in the charged
\tHp\ channel, the HO and spherical-wall extractions remain in good agreement
with the continuum result; a periodic-box Coulomb calculation lies outside the
scope of the present study. A Monte Carlo analysis further shows that spectral
errors are amplified near trap-function poles and along the poorly conditioned
inelasticity and phase-shift-difference directions, whereas the phase-shift sum
is comparatively robust. The periodic-box extraction is performed in a
single-partial-wave truncation, with higher partial waves subduced into the
same cubic irreducible representation neglected.

The paper is organized as follows. Section~\ref{sec:theory} defines the
coupled-channel Hamiltonian, scattering conventions, and calculable
$R$-matrix benchmark. Section~\ref{sec:qcs} presents the QCs for the three
confinement geometries in a common notation. Section~\ref{sec:extraction}
describes the overdetermined extraction procedure. Numerical results and the
propagation of spectral uncertainties are presented in
Sec.~\ref{sec:results}. Section~\ref{sec:conclusion} summarizes our
conclusions.

%==================================
\section{Coupled-channel framework}
\label{sec:theory}
%==================================

We label the \tHp\ and \tHen\ channels by $c=1$ and $c=2$, respectively,
and denote their thresholds by $E_1$ and $E_2$. The total energy is measured
relative to the \tHp\ threshold, so that
\begin{equation}
    E_1=0,
    \qquad
    E_2-E_1=0.763~\MeV.
    \label{eq:thresholds}
\end{equation}
For fixed orbital angular momentum $\ell$, channel spin, and total angular
momentum, the radial coupled-channel Hamiltonian is
\begin{equation}
    H=
    \begin{pmatrix}
        T_1+E_1+V_d(r)+V_C(r) & V_o(r) \\
        V_o(r)                 & T_2+E_2+V_d(r)
    \end{pmatrix},
    \label{eq:Hamiltonian}
\end{equation}
where
\begin{equation}
    T_c=-\frac{\hbar^2}{2\mu_c}
    \left[
        \frac{d^2}{dr^2}-\frac{\ell(\ell+1)}{r^2}
    \right]
    \label{eq:radial_kinetic_energy}
\end{equation}
is the radial kinetic-energy operator in channel $c$, and $\mu_c$ is the
corresponding reduced mass. We take the reduced masses in the two channels to
be equal,
\begin{equation}
    \mu_1=\mu_2=\frac{3}{4}m_N,
    \qquad
    m_N=938.918~\MeV.
    \label{eq:reduced_masses}
\end{equation}

The coupled radial wave function is
\begin{equation}
    \bm{u}(r)=
    \begin{pmatrix}
        u_1(r) \\
        u_2(r)
    \end{pmatrix}.
    \label{eq:coupled_wave_function}
\end{equation}

Following Refs.~\cite{Suzuki2008,Odsuren2021,ZhangCoupled2024}, the nuclear
interaction is expressed in terms of its $T=1$ and $T=0$ isospin components.
The diagonal and off-diagonal potentials are
\begin{align}
    V_d(r)
    &=\frac{1}{2}\left[
        V_1\exp\left(-\frac{r^2}{b_1^2}\right)
        +V_0\exp\left(-\frac{r^2}{b_0^2}\right)
    \right],
    \label{eq:diagonal_potential}
    \\
    V_o(r)
    &=\frac{1}{2}\left[
        V_1\exp\left(-\frac{r^2}{b_1^2}\right)
        -V_0\exp\left(-\frac{r^2}{b_0^2}\right)
    \right],
    \label{eq:coupling_potential}
\end{align}
where $V_1\equiv V_{T=1}$ and $V_0\equiv V_{T=0}$ are the interaction
strengths, and $b_1$ and $b_0$ are the corresponding Gaussian ranges. We
consider the $^1S_0$ and $^3P_1$ partial waves. The parameters used in the
calculations are listed in Table~\ref{tab:interaction_parameters}.

\begin{table}[t]
    \caption{Parameters of the Gaussian nuclear interaction for the $^1S_0$
    and $^3P_1$ partial waves. The strengths $V_1$ and $V_0$ are given in
    \MeV, and the ranges $b_1$ and $b_0$ in \fm.}
    \label{tab:interaction_parameters}
    \begin{ruledtabular}
        \begin{tabular}{ccccc}
            Partial wave & $V_1$ & $b_1$ & $V_0$ & $b_0$ \\
            \hline
            $^1S_0$ & $-27.60$ & $3.00$ & $-58.50$ & $3.00$ \\
            $^3P_1$ & $-18.83$ & $3.06$ & $-8.00$  & $3.00$
        \end{tabular}
    \end{ruledtabular}
\end{table}

The charged \tHp\ channel also contains the regularized finite-size Coulomb
interaction
\begin{equation}
    V_C(r)=\frac{e^2}{r}
    \operatorname{erf}\!\left(\sqrt{\beta}\,r\right),
    \label{eq:coulomb_erf}
\end{equation}
with $e^2=1.44~\MeV\,\fm$ and $\beta=0.66~\fm^{-2}$. The
error-function regularization removes the point-Coulomb singularity at the
origin while preserving the asymptotic behavior $V_C(r)\to e^2/r$ at large
distances.

For an open channel, we define the kinetic energy relative to threshold and
the corresponding momentum by
\begin{equation}
    \varepsilon_c(E)=E-E_c,
    \qquad
    k_c(E)=\frac{\sqrt{2\mu_c\varepsilon_c(E)}}{\hbar}.
    \label{eq:channel_kinematics}
\end{equation}

For two open channels, time-reversal invariance and unitarity allow the
scattering matrix to be parametrized as
\begin{equation}
    \Smat(E)=
    \begin{pmatrix}
        \eta(E)e^{2i\delta_1(E)} & \xi(E) \\
        \xi(E) & \eta(E)e^{2i\delta_2(E)}
    \end{pmatrix},
    \label{eq:S_parameterization}
\end{equation}
where
\begin{equation}
    \xi(E)=i\sqrt{1-\eta^2(E)}\,
    e^{i[\delta_1(E)+\delta_2(E)]}.
    \label{eq:S_off_diagonal}
\end{equation}
Here, $\delta_1(E)$ and $\delta_2(E)$ are the phase shifts in the \tHp\ and
\tHen\ channels, respectively, and $0\leq\eta(E)\leq1$ is the inelasticity.
When Coulomb interactions are present, $\delta_1$ is defined relative to the
Coulomb functions introduced below. The elastic limit corresponds to
$\eta=1$, whereas $\eta<1$ indicates flux transfer between the two open
channels. We define the dimensionless reaction matrix $\Kmat$ by
\begin{equation*}
\begin{aligned}
    \Kmat(E)
    &=-i\left[\Smat(E)-\bm{1}\right]
      \left[\Smat(E)+\bm{1}\right]^{-1},
    \\
    \Smat(E)
    &=[\bm{1}+i\Kmat(E)][\bm{1}-i\Kmat(E)]^{-1}.
\end{aligned}
\end{equation*}
For a time-reversal-invariant system, $\Kmat$ is real and symmetric. In a
single uncoupled channel, this convention gives $\Kmat=\tan\delta$.

Below the \tHen\ threshold, only the \tHp\ channel is asymptotically open,
and the observable scattering matrix is characterized by a single elastic
phase shift.

%----------------------
\subsection{$R$-matrix}
\label{sec:Rmatrix}
%----------------------

We use the multichannel $R$-matrix method to obtain continuum reference
results for the confinement extractions. The formalism partitions the relative
radial coordinate at a channel radius $a$ into internal and external regions.
The channel radius is chosen sufficiently large that the short-range nuclear
interaction and channel coupling are negligible for $r>a$. In the internal
region, $H^{(B)}$ denotes the radial coupled-channel Hamiltonian of
Eq.~\eqref{eq:Hamiltonian}, restricted to $0<r<a$ and acting on wave functions
that are regular at the origin and satisfy auxiliary logarithmic-derivative
boundary conditions at $r=a$. Its eigenvalue problem is
\begin{equation}
 H^{(B)}\bm u^{(B)}_\lambda=E^{(B)}_\lambda\bm u^{(B)}_\lambda,
 \qquad
 a u^{(B)\prime}_{\lambda c}(a)=B_c u^{(B)}_{\lambda c}(a).
\label{eq:R_formal_basis}
\end{equation}
The real constants $B_c$ render $H^{(B)}$ self-adjoint. The corresponding
eigenfunctions form a discrete basis for constructing the physical internal
scattering solution, which is then matched at $r=a$ to the known external
Coulomb and free-particle solutions. These eigenfunctions are formal basis
states rather than physical scattering states, and their eigenvalues
$E^{(B)}_\lambda$ are formal pole energies rather than physical resonance
energies. Although these formal quantities depend on the choice of $B_c$, the
resulting scattering matrix does not. In the numerical calculations, we set
$B_1=B_2=1$. For notational simplicity, we henceforth suppress the superscript
$(B)$ on the formal energies and basis functions.

At a fixed scattering energy $E$, the internal part of a physical solution can
be expanded as
\begin{equation}
 u_c(r;E)=\sum_\lambda \alpha_\lambda(E)u_{\lambda c}(r),
 \qquad 0<r<a.
\label{eq:R_internal_expansion}
\end{equation}
Applying Green's identity to the physical solution and a formal basis state
yields
\begin{equation}
 \alpha_\lambda(E)=\frac{1}{E_\lambda-E}\sum_c
 \gamma_{\lambda c}\left(\mathcal{D}_c-B_c\mathcal{V}_c\right),
\label{eq:R_expansion_coefficients}
\end{equation}
where $\gamma_{\lambda c}$ is determined by the surface value of the formal
basis state, and $\mathcal{V}_c$ and $\mathcal{D}_c$ encode the surface value
and derivative of the physical scattering solution:
\begin{align}
 \gamma_{\lambda c}
 &=\left(\frac{\hbar^2}{2\mu_c a}\right)^{1/2}u_{\lambda c}(a),
 \label{eq:R_surface_data_formal}
 \\
 \mathcal{V}_c
 &=\left(\frac{\hbar^2}{2\mu_c a}\right)^{1/2}u_c(a),
 \qquad
 \mathcal{D}_c
 =\left(\frac{\hbar^2}{2\mu_c a}\right)^{1/2}a u'_c(a).
\label{eq:R_surface_data_physical}
\end{align}
Evaluating Eq.~\eqref{eq:R_internal_expansion} at $r=a$ gives the boundary
relation
\begin{equation}
\mathcal{V}_c
=
\sum_{c'} R_{cc'}(E)
\left(
\mathcal{D}_{c'}-B_{c'}\mathcal{V}_{c'}
\right),
\label{eq:R_matrix_boundary_relation}
\end{equation}
with
\begin{equation}
R_{cc'}(E)
=
\sum_\lambda
\frac{\gamma_{\lambda c}\gamma_{\lambda c'}}
{E_\lambda-E}.
\label{eq:R_matrix_definition}
\end{equation}

For $r>a$ and incident-channel amplitudes $y_{c'}$, the external radial
solution in channel $c$ can be written as
\begin{equation}
\begin{aligned}
    u_c(r;E)
    ={}&\sqrt{\frac{\mu_c}{k_c}}
    \sum_{c'}
    \bigl[\delta_{cc'} I_c(k_c r)
    \\
    &\hspace{3.6em}- {\Smat}_{cc'} O_c(k_c r)\bigr]y_{c'}.
\end{aligned}
    \label{eq:ext_sol}
\end{equation}
For the charged channel, $I_c$ and $O_c$ are the incoming and outgoing
Coulomb functions,
\begin{equation}
\begin{aligned}
O_c(\rho)
&=
G_{\ell}(\eta_{C,c},\rho)
+iF_{\ell}(\eta_{C,c},\rho),
\\
I_c(\rho)
&=
G_{\ell}(\eta_{C,c},\rho)
-iF_{\ell}(\eta_{C,c},\rho),
\end{aligned}
\label{eq:R_coulomb_functions}
\end{equation}
where the Sommerfeld parameter is
\begin{equation}
\eta_{C,c}(E)
=
\frac{Z_{1c}Z_{2c}\mu_c e^2}
{\hbar^2 k_c(E)}.
\label{eq:R_sommerfeld_parameter}
\end{equation}
At fixed $E$, the channel dependence of the Sommerfeld parameter in the
one-argument functions $I_c$ and $O_c$ is implicit.
For the neutral channel, $\eta_{C,c}=0$, and these functions reduce to the
corresponding free-particle solutions. The Sommerfeld parameter $\eta_{C,c}$
should not be confused with the inelasticity $\eta(E)$ introduced in
Eq.~\eqref{eq:S_parameterization}.

At the channel radius, we define the dimensionless coordinate
\begin{equation}
\rho_c(E)=k_c(E)a.
\label{eq:R_external_variables}
\end{equation}
We then use the shorthand
\begin{equation}
\begin{aligned}
I_c(E)&\equiv I_c\!\left(\rho_c(E)\right),
\\
O_c(E)&\equiv O_c\!\left(\rho_c(E)\right).
\end{aligned}
\label{eq:R_asymptotic_shorthand}
\end{equation}
These are the incoming and outgoing functions evaluated at $r=a$. Derivatives
with respect to the dimensionless radial coordinate, evaluated at $\rho_c(E)$,
are denoted by $I'_c(E)$ and $O'_c(E)$.

Applying Eq.~\eqref{eq:R_matrix_boundary_relation} to the external solution
in Eq.~\eqref{eq:ext_sol} yields the relation
\begin{equation}
 \Smat(E)
 =
 \left[\bm M(E)\bm O(E)-\bm O'(E)\right]^{-1}
 \left[\bm M(E)\bm I(E)-\bm I'(E)\right],
 \label{eq:R_collision_matrix_M}
\end{equation}
where
\begin{equation}
 \bm{\Lambda}(E)=\bm B+\bm R^{-1}(E),
 \qquad
 \bm M(E)=\bm\rho^{-1/2}\bm{\Lambda}(E)\bm\rho^{-1/2}.
 \label{eq:R_M_definitions}
\end{equation}
The channel-space matrices are
\begin{equation*}
\begin{aligned}
\bm I&=\operatorname{diag}(I_c), &
\bm O&=\operatorname{diag}(O_c),
\\
\bm\rho&=\operatorname{diag}(\rho_c), &
\bm B&=\operatorname{diag}(B_c).
\end{aligned}
\end{equation*}
Defining
\begin{equation}
 \bm{\mathcal L}
 = \bm\rho \bm O' \bm O^{-1},
 \qquad
 \bm{\mathcal L}_0
 =
 \bm{\mathcal L}-\bm B,
 \label{eq:R_log_derivative}
\end{equation}
Eq.~\eqref{eq:R_collision_matrix_M} can equivalently be written in the
standard $R$-matrix form
\begin{equation}
 \Smat
 =
 \bm I\bm O^{-1}
 +
 2i\,\bm\rho^{1/2}\bm O^{-1}
 \left(\bm{1}-\bm R\bm{\mathcal L}_0\right)^{-1}
 \bm R\,
 \bm\rho^{1/2}\bm O^{-1}.
 \label{eq:R_collision_matrix_standard}
\end{equation}
The phase shifts and inelasticity are obtained by matching this matrix to
Eq.~\eqref{eq:S_parameterization}.

To construct the $R$-matrix reference numerically, we discretize the internal
eigenvalue problem in Eq.~\eqref{eq:R_formal_basis} on a uniform radial grid.
The interval $0<r<a$, with $a=9~\fm$, is divided into $N=300$ steps of size
$\Delta r=a/N=0.03~\fm$, yielding $299$ interior grid points per channel.
Using the logarithmic-derivative boundary condition in
Eq.~\eqref{eq:R_formal_basis}, the surface value entering
Eq.~\eqref{eq:R_surface_data_formal} is approximated by
\begin{equation}
u_{\lambda c}(a) =
\frac{u_{\lambda c}(a-\Delta r)}
     {1-B_c\Delta r/a}.
\label{eq:R_surface_reconstruction}
\end{equation}

%===========================================================
\section{Artificial confinement and quantization conditions}
\label{sec:qcs}
%===========================================================

Artificial confinement is introduced by adding a known external operator
$U_{\rm trap}$ to the coupled-channel Hamiltonian. The confined eigenvalue
problem is
\begin{equation}
    \left[H+U_{\rm trap}\right]
    \bm{u}_{n}^{\rm trap}(r)
    =E_{n}^{\rm trap}\bm{u}_{n}^{\rm trap}(r).
    \label{eq:trapped_eigenvalue_problem}
\end{equation}
All energies entering the QCs refer to relative motion; any center-of-mass
contribution is removed before the extraction.

For each geometry, we define a dimensionless channel-dependent trap function
$\mathcal{F}_c(E;\lambda)$, where $\lambda$ denotes the oscillator frequency
$\omega$, spherical-wall radius $R$, or periodic-box length $L$. The same
confinement parameter is applied to both channels, but the corresponding trap
functions differ because the channel thresholds, momenta, and Coulomb
interactions differ. In a single-channel problem with one retained partial
wave, our convention is
\begin{equation}
    \cot\delta_c(E)-\mathcal{F}_c(E;\lambda)=0.
    \label{eq:single_channel_schematic}
\end{equation}
Thus, $\mathcal{F}_c$ contains the geometric and kinematic dependence of the
confinement, whereas $\delta_c$ contains the infinite-volume scattering
information.

For two coupled open channels, with one partial wave retained in each channel,
the QC is~\cite{GuoLong2022,ZhangCoupled2024}
\begin{equation}
\begin{aligned}
    0={}&
    \eta(1+\mathcal{F}_1\mathcal{F}_2)
    \cos(\delta_1-\delta_2)
    \\
    &+(1-\mathcal{F}_1\mathcal{F}_2)
    \cos(\delta_1+\delta_2)
    \\
    &-\eta(\mathcal{F}_1-\mathcal{F}_2)
    \sin(\delta_1-\delta_2)
    \\
    &-(\mathcal{F}_1+\mathcal{F}_2)
    \sin(\delta_1+\delta_2).
\end{aligned}
    \label{eq:two_channel_QC}
\end{equation}
Here, $\delta_c=\delta_c(E)$, $\eta=\eta(E)$, and
$\mathcal{F}_c=\mathcal{F}_c(E;\lambda)$. Equation~\eqref{eq:two_channel_QC}
is the central relation used in the extraction; the confinement geometry
enters only through the functions $\mathcal{F}_c$.

Using the reaction matrix defined above, the same condition has the compact
determinant form
\begin{equation*}
    \det\!\left[
        \Kmat^{-1}(E)
        -\operatorname{diag}\!\left(\mathcal{F}_1,\mathcal{F}_2\right)
    \right]=0.
\end{equation*}
Expanding this determinant with the parametrization in
Eqs.~\eqref{eq:S_parameterization} and~\eqref{eq:S_off_diagonal} gives
Eq.~\eqref{eq:two_channel_QC}.

%------------------------------------
\subsection{Harmonic-oscillator trap}
%------------------------------------

For a common oscillator frequency $\omega$, the HO confinement operator is
\begin{equation}
    U_{\rm HO}(r)=
    \begin{pmatrix}
        \dfrac{1}{2}\mu_1\omega^2r^2 & 0 \\
        0 & \dfrac{1}{2}\mu_2\omega^2r^2
    \end{pmatrix}.
    \label{eq:HO_trap}
\end{equation}
For a neutral channel, the Busch--Englert--Rza\.{z}ewski--Wilkens (BERW)
trap function for partial wave $\ell$ is
\begin{equation}
\begin{aligned}
    \mathcal{F}_{\ell,c}^{\rm HO}(E;\omega)
    ={}&(-1)^{\ell+1}
    \left(\frac{4\mu_c\omega}{\hbar k_c^2}\right)^{\ell+1/2}
    \\
    &\times
    \frac{\Gamma\!\left(\dfrac{3}{4}+\dfrac{\ell}{2}
            -\dfrac{\varepsilon_c(E)}{2\hbar\omega}\right)}
         {\Gamma\!\left(\dfrac{1}{4}-\dfrac{\ell}{2}
            -\dfrac{\varepsilon_c(E)}{2\hbar\omega}\right)}.
\end{aligned}
    \label{eq:berw}
\end{equation}
The corresponding oscillator length is
$b_c=\sqrt{\hbar/(\mu_c\omega)}$. The asymptotic QC applies when $b_c$ is
large compared with the range of the nuclear interaction; residual
finite-range trap effects vanish in the weak-confinement limit.

For the charged channel, Eq.~\eqref{eq:berw} is replaced by a
Coulomb-modified trap function. In the Green-function representation used
here, it can be written as
\begin{equation}
\begin{aligned}
    \mathcal{F}_{\ell,c}^{\rm HO,C}(E;\omega)
    ={}&
    \frac{\hbar^2}
         {2\mu_c k_c^{2\ell+1}C_\ell^2(\eta_{C,c})}
    \\
    &\times
    \left.
    \frac{\Delta G_{\ell,c}^{C}(r,r';E,\omega)}{(rr')^\ell}
    \right|_{r,r'\to0},
\end{aligned}
    \label{eq:coulomb_F_symbolic}
\end{equation}
where $C_\ell(\eta_{C,c})$ is the generalized Sommerfeld factor and
$\Delta G_{\ell,c}^{C}$ is the regularized difference between the
infinite-volume and trapped Coulomb Green functions. The sign convention is
chosen so that Eq.~\eqref{eq:berw} is recovered in the neutral limit. The
required Green function can be evaluated perturbatively or by solving the
corresponding Dyson equation numerically
\cite{Guo2021,ZhangCharged2024,Bagnarol2025}.
The direct-inversion implementation of Ref.~\cite{Bagnarol2025}, adopted here,
is presently available for $\ell=0$ and $1$, the two partial waves considered
in this work.

The finite-size Coulomb interaction in Eq.~\eqref{eq:coulomb_erf} has the same
long-range behavior as the point-Coulomb potential. Its short-range difference
from $e^2/r$ remains part of the interaction Hamiltonian, while the long-range
Coulomb behavior is incorporated into the trap function. The noninteracting
trapped Green function has poles at
\begin{equation}
    \frac{\varepsilon_c(E)}{\hbar\omega}
    =2n+\ell+\frac{3}{2},
    \qquad n=0,1,2,\ldots.
    \label{eq:HO_noninteracting_poles}
\end{equation}
The direct Coulomb Green-function calculation remains stable in neighborhoods
on either side of these poles; the pole positions themselves are analytic
singularities of the trap function.

%------------------------------------
\subsection{Spherical hard-wall trap}
%------------------------------------

For a spherical hard wall of radius $R$, each channel component satisfies
\begin{equation}
    u_c(R)=0.
    \label{eq:spherical_wall_boundary}
\end{equation}
For a neutral channel, the corresponding trap function is
\begin{equation}
    \mathcal{F}_{\ell,c}^{\rm SW}(E;R)
    =\frac{n_\ell(k_cR)}{j_\ell(k_cR)},
    \label{eq:spherical_F}
\end{equation}
where $j_\ell$ and $n_\ell$ are the spherical Bessel functions of the first
and second kind, respectively.

For a charged channel, the free asymptotic functions are replaced by their
Coulomb-modified counterparts. Using the same regular and irregular Coulomb
functions as in Eq.~\eqref{eq:R_coulomb_functions}, the trap function
is~\cite{Guo2021}
\begin{equation}
    \mathcal{F}_{\ell,c}^{\rm SW,C}(E;R)
    =-\frac{G_\ell(\eta_{C,c},k_cR)}
            {F_\ell(\eta_{C,c},k_cR)}.
    \label{eq:spherical_coulomb_F}
\end{equation}
The neutral limits of the Coulomb functions are
\begin{equation}
\begin{aligned}
    \lim_{\eta_{C,c}\to0}
    F_\ell(\eta_{C,c},k_cR)
    &=k_cR\,j_\ell(k_cR),
    \\
    \lim_{\eta_{C,c}\to0}
    G_\ell(\eta_{C,c},k_cR)
    &=-k_cR\,n_\ell(k_cR),
\end{aligned}
    \label{eq:spherical_coulomb_neutral_limit}
\end{equation}
so Eq.~\eqref{eq:spherical_coulomb_F} reduces continuously to
Eq.~\eqref{eq:spherical_F} in the neutral limit. The spherical wall is
particularly convenient for the charged channel because the Coulomb
contribution is incorporated to all orders through the asymptotic Coulomb
functions.

%------------------------------
\subsection{Periodic cubic box}
%------------------------------

For a cubic volume of side length $L$ with periodic boundary conditions,
continuous rotational symmetry, $\mathrm{SO}(3)$, is reduced to the cubic
group $O_h$. States are therefore classified by irreducible representations
(irreps) of $O_h$ rather than by orbital angular momentum. Each cubic irrep
contains infinitely many partial waves, although higher-partial-wave contributions
are generally suppressed at sufficiently low energies
\cite{Johnson1982,LuuSavage2011}.

For channel $c$, we define
\begin{equation}
    q_c=\frac{k_cL}{2\pi}.
    \label{eq:dimensionless_box_momentum}
\end{equation}
In the $A_1^+$ irrep, neglecting contributions from $\ell=4,6,\ldots$, the
$s$-wave trap function is
\begin{equation}
    \mathcal{F}_{0,c}^{\rm box}(E;L)
    =\frac{\mathcal{Z}_{00}(1;q_c^2)}{\pi^{3/2}q_c}
    =\frac{2}{\sqrt{\pi}\,k_cL}
     \mathcal{Z}_{00}(1;q_c^2),
    \label{eq:luscher_swave}
\end{equation}
where $\mathcal{Z}_{00}$ is the analytically continued L\"uscher zeta
function. In the $T_1^-$ irrep, within the truncation in which the
$\ell=3,5,\ldots$ phase shifts are neglected, the $p$-wave QC has the same
scalar trap function,
\begin{equation}
    \mathcal{F}_{1,c}^{\rm box}(E;L)
    =\frac{\mathcal{Z}_{00}(1;q_c^2)}{\pi^{3/2}q_c}.
    \label{eq:luscher_pwave}
\end{equation}
This equality applies only within the single-partial-wave truncation. Once
higher partial waves belonging to the same irrep are retained, the
finite-volume condition becomes a matrix equation.

More generally, including both reaction channels and cubic partial-wave
mixing, the QC can be written as
\begin{equation}
    \det\!\left[
        \widetilde{\Kmat}^{-1}(E)-B^{\Gamma}(E,L)
    \right]=0,
    \label{eq:box_general}
\end{equation}
where $\widetilde{\Kmat}$ is a threshold-rescaled form of the infinite-volume
$K$ matrix defined above, and $B^{\Gamma}$ is the known finite-volume box
matrix for the cubic irrep $\Gamma$~\cite{Guo2013,Briceno2014,Morningstar2017}.
The two matrices must be expressed in the same normalization. For example, if
$\bm N$ is diagonal in channel and partial-wave space, with
$N_{c\ell,c'\ell'}=\delta_{cc'}\delta_{\ell\ell'}k_c^{\ell+1/2}$, one may
choose $\widetilde{\Kmat}^{-1}=\bm N\Kmat^{-1}\bm N$ and absorb the corresponding
powers of $2\pi/L$ into $B^{\Gamma}$. The determinant acts in channel space,
in the space of partial waves subduced into $\Gamma$, and in any additional
spin spaces required by the system.

When partial-wave mixing is neglected, Eq.~\eqref{eq:two_channel_QC} may be
used with the scalar functions $\mathcal{F}_{\ell,c}^{\rm box}$. For the
low-energy $^1S_0$ calculation, the leading omitted contribution is the
$\ell=4$ wave in the $A_1^+$ irrep; for the $^3P_1$ calculation, it is the
$\ell=3$ wave in the $T_1^-$ irrep. Their effects are assumed to be small over
the energy range considered, but they are not isolated in the present
comparison.

%============================
\section{Extraction strategy}
\label{sec:extraction}
%============================

Above the second-channel threshold, the scattering matrix is characterized at
each energy by two phase shifts and one inelasticity. We collect these
observables in the parameter vector
\begin{equation}
    \bm{x}(E)=
    \begin{pmatrix}
        \delta_1(E) \\
        \delta_2(E) \\
        \eta(E)
    \end{pmatrix}.
    \label{eq:scattering_parameter_vector}
\end{equation}
For a confined level at energy $E$, the coupled-channel QC provides one
nonlinear constraint,
\begin{equation}
    q_a(\bm{x};E)
    \equiv
    \mathcal{Q}\!\left[\bm{x}(E);E,\lambda_a\right]
    =0,
    \label{eq:single_trap_constraint}
\end{equation}
where $\mathcal{Q}$ denotes the right-hand side of
Eq.~\eqref{eq:two_channel_QC}, and $\lambda_a$ is the confinement parameter
associated with the $a$th constraint. Depending on the geometry, $\lambda_a$
is an oscillator frequency $\omega_a$, a spherical-wall radius $R_a$, or a
periodic-box length $L_a$.

A single level therefore supplies only one relation among the three unknown
observables. In principle, at least three independent constraints, obtained
from distinct levels or confinement parameters that cross the same target
energy, are required to determine $\delta_1(E)$, $\delta_2(E)$, and $\eta(E)$.
In practice, additional constraints improve the stability of the inversion.
We assemble the residuals into
$\bm q=(q_1,\ldots,q_{N_{\rm con}})^{\mathsf T}$ and perform the
overdetermined generalized least-squares fit
\begin{equation}
    \widehat{\bm{x}}(E)
    =\operatorname*{arg\,min}_{\delta_1,\delta_2,\,0\leq\eta\leq1}
    \chi^2(\delta_1,\delta_2,\eta;E),
    \label{eq:least_squares_definition}
\end{equation}
with
\begin{equation}
    \chi^2(\delta_1,\delta_2,\eta;E)
    =\bm q^{\mathsf T}\bm W\bm q.
    \label{eq:least_squares}
\end{equation}
Here, $N_{\rm con}$ is the number of constraints and $\bm W$ is a
positive-definite weight matrix. We set $\bm W=\bm{1}$ for the central
benchmark calculations. In the uncertainty analysis of
Sec.~\ref{sec:sensitivity}, $\bm W$ is obtained from the estimated covariance
of the QC residuals.

At a target energy $E$, the extraction proceeds as follows:
\begin{enumerate}
    \item Choose a confinement geometry and target scattering energy.

    \item Compute several confined energy levels over a range of confinement
    parameters $\lambda$.

    \item Represent each selected level as a function of $\lambda$ using a
    shape-preserving piecewise cubic Hermite interpolant (PCHIP), and determine
    the values $\lambda_a$ at which the level crosses the target energy.

    \item Evaluate $\mathcal{F}_1(E;\lambda_a)$ and
    $\mathcal{F}_2(E;\lambda_a)$ for each resulting constraint.

    \item Extract $\delta_1(E)$, $\delta_2(E)$, and $\eta(E)$ by minimizing
    Eq.~\eqref{eq:least_squares}, subject to $0\leq\eta\leq1$.

    \item Repeat the procedure for each energy on the selected grid.
\end{enumerate}

After completing the extraction, we compare the resulting observables with
the calculable $R$-matrix benchmark described in Sec.~\ref{sec:Rmatrix}. This
comparison provides a direct validation of the confinement-based inversion but
is not itself part of the extraction procedure.

Below the \tHen\ threshold, only the \tHp\ channel is asymptotically open. The
observable scattering matrix is then characterized by a single elastic phase
shift, and the corresponding single-channel QC is used, with its
Coulomb-modified form employed when the Coulomb interaction is included.
Coupling to the closed \tHen\ channel remains included in the confined
Hamiltonian. Above threshold, the full two-channel QC in
Eq.~\eqref{eq:two_channel_QC} is required.

%==================
\section{Results}
\label{sec:results}
%==================

%---------------------------------------------------------
\subsection{Confined spectra and numerical implementation}
%---------------------------------------------------------

We extract the coupled-channel observables from spectra generated by the HO
trap, spherical hard wall, and periodic cubic box. The HO spectra are computed
on a uniform grid of $81$ frequencies over $\hbar\omega\in[0.1,0.5]~\MeV$, with
$\Delta(\hbar\omega)=0.005~\MeV$. The spherical-wall spectra use
$R/\fm=10,11,\ldots,50$ for the $^1S_0$ partial wave and
$R/\fm=40,41,\ldots,120$ for the $^3P_1$ partial wave. The periodic-box
spectra are computed on a uniform grid over $L\in[10,80]~\fm$, with
$\Delta L=0.2~\fm$. The central extractions in
Figs.~\ref{fig:results_1s0} and~\ref{fig:results_3p1} use all available
crossings at each energy. For the uncertainty comparison in
Sec.~\ref{sec:sensitivity}, the number of constraints is matched across the
three geometries.

The HO and spherical-wall eigenvalue problems are solved in radial coordinates
with the matrix Numerov method~\cite{Pillai2012,Barnea2026} applied to the
coupled-channel Hamiltonian.
For the periodic box, the Hamiltonian is discretized on a three-dimensional
Cartesian grid with periodic boundary conditions, using a finite-difference
implementation based on the algorithm of Ref.~\cite{Konig2020}.
We extend this implementation to project onto the relevant cubic irreps:
$A_1^+$ for the $s$ wave and $T_1^-$ for the $p$ wave.

For each discrete level, a PCHIP interpolant is constructed as a function of
the relevant confinement parameter, following Sec.~\ref{sec:extraction}. The
interpolants locate the values of $\omega$, $R$, or $L$ at which independent
levels cross a selected scattering energy. The channel-dependent trap
functions are evaluated at these crossings and inserted into the
coupled-channel QC, and the resulting overdetermined residual vector is fitted
with Eq.~\eqref{eq:least_squares}.

\begin{figure}[t]
\centering
\includegraphics[width=\columnwidth]{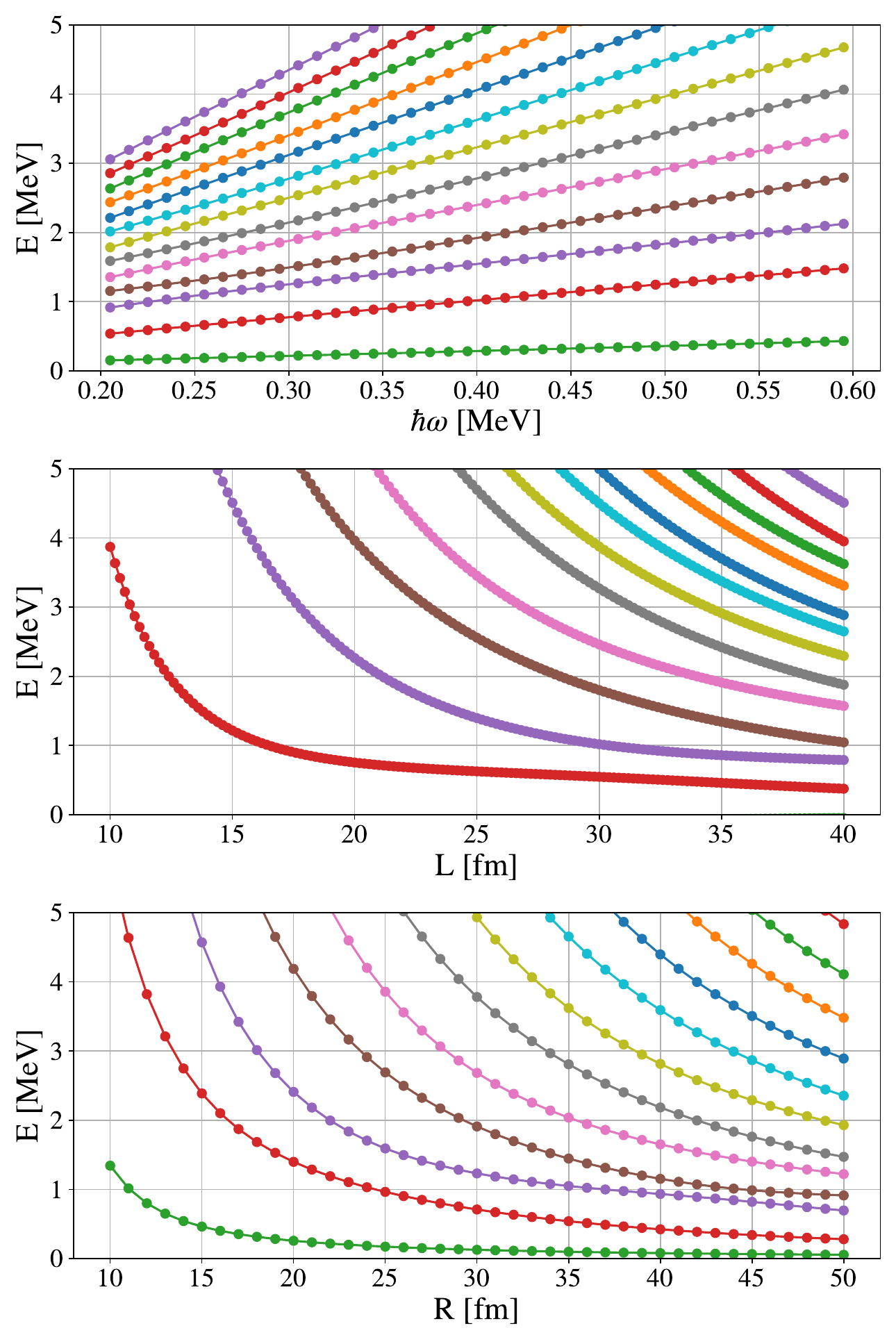}
\caption{Coupled-channel $^1S_0$ spectra for the three confinement
geometries. From top to bottom, the panels show the discrete energies as
functions of the oscillator energy $\hbar\omega$, periodic-box length $L$, and
spherical-wall radius $R$. Each curve represents one confined level. PCHIP
interpolation of the individual curves locates the confinement parameters at
which several levels cross a common scattering energy.}
\label{fig:energy_levels}
\end{figure}

Figure~\ref{fig:energy_levels} shows representative confined $^1S_0$ spectra.
The HO energies increase approximately linearly with $\hbar\omega$, whereas
the positive-energy levels in the periodic box and spherical hard wall
decrease as $L$ or $R$ increases. The latter behavior reflects the weakening
confinement and increasing density of states as the continuum limit is
approached.

%---------------------------
\subsection{$^1S_0$ results}
%---------------------------

\begin{figure*}[t]
\centering
\includegraphics[width=0.49\textwidth]{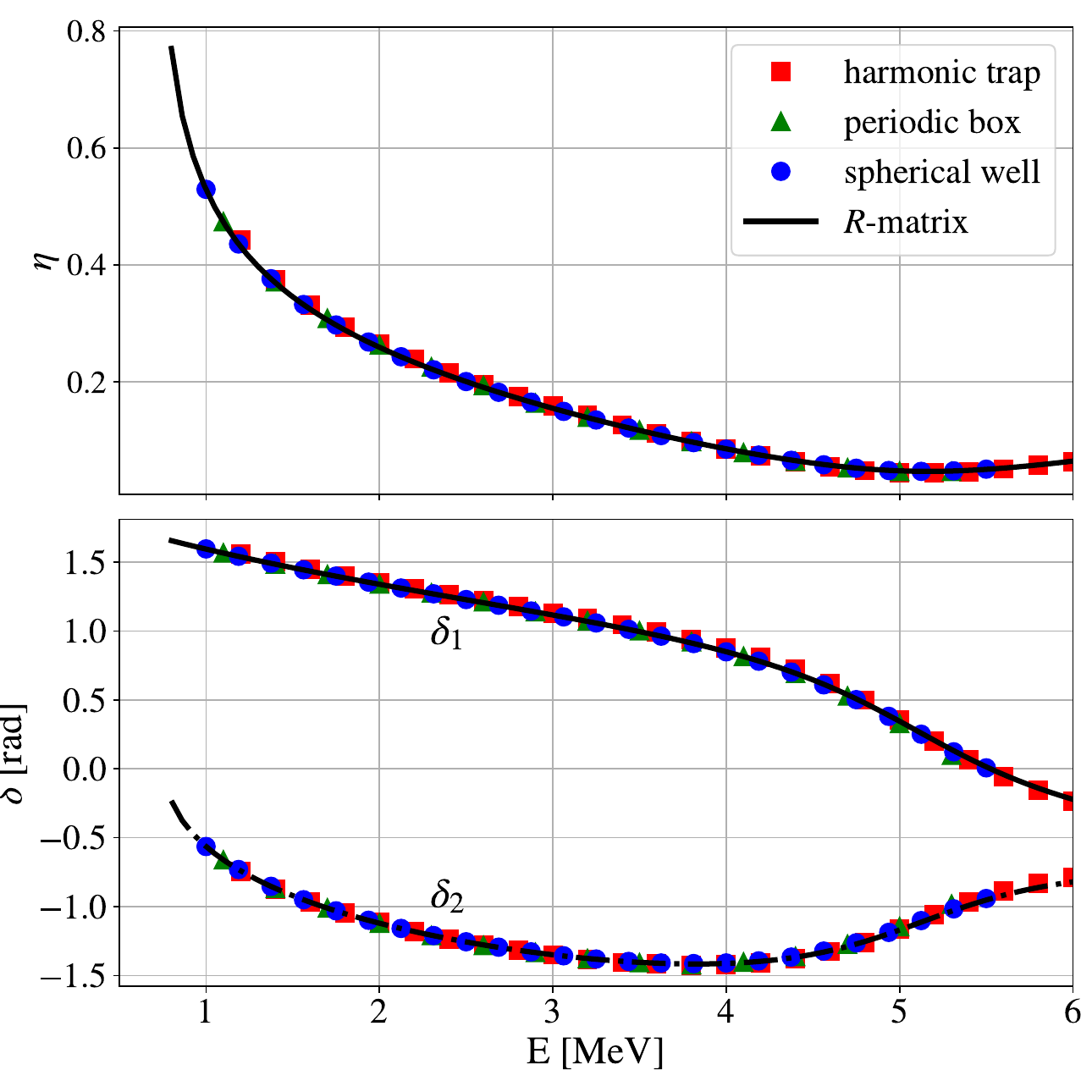}
\hfill
\includegraphics[width=0.49\textwidth]{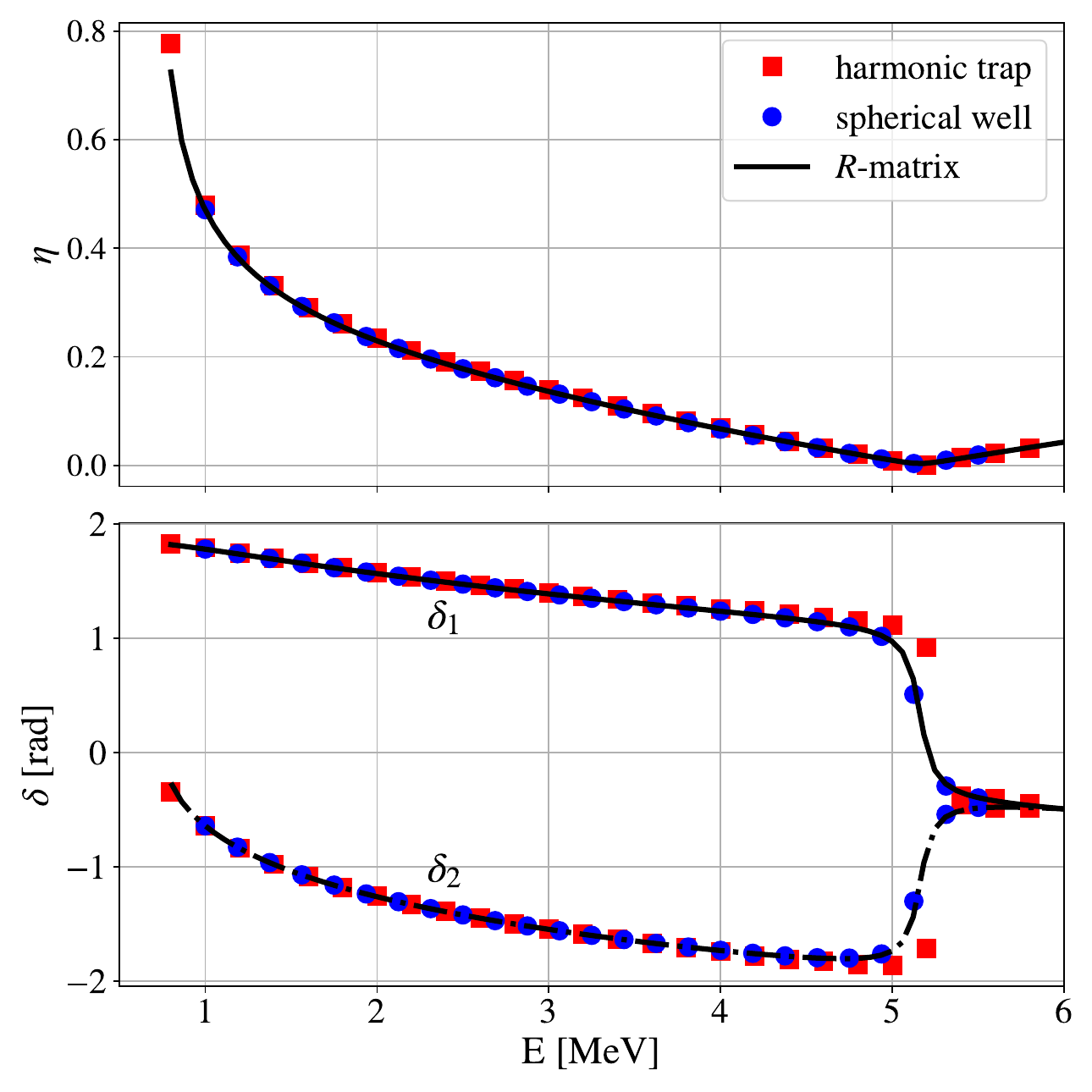}
\caption{Inelasticity and phase shifts in the $^1S_0$ partial wave as functions of
scattering energy. The left panel omits the Coulomb interaction, while the
right panel includes it in the charged \tHp\ channel. In each panel, the upper
plot shows $\eta(E)$ and the lower plot shows $\delta_1(E)$ and
$\delta_2(E)$. Black curves are the $R$-matrix reference; colored symbols are
the observables extracted from the HO trap (red), spherical hard wall (blue),
and periodic cubic box (green). Periodic-box results are not included in the
calculation with Coulomb interactions.}
\label{fig:results_1s0}
\end{figure*}

Figure~\ref{fig:results_1s0} presents the extracted $^1S_0$ observables.
Without Coulomb interactions, all three confinement schemes closely reproduce
the continuum reference throughout the displayed energy interval. Above the
second-channel threshold, the extracted inelasticity follows the $R$-matrix
curve through its decrease to an intermediate-energy minimum and its
subsequent mild rise. The three schemes also reproduce the gradual variation
of $\delta_1$ and the nonmonotonic energy dependence of $\delta_2$.

Small differences become visible only toward the upper end of the interval,
where the extraction is more sensitive to the numerical precision of the
confined spectra, the interpolation to common scattering energies, and the
conditioning of the nonlinear inverse problem. These differences remain small
and do not alter the overall consistency among the three geometries.

With Coulomb interactions included, the HO and spherical-wall extractions
again agree closely with the $R$-matrix reference. The inelasticity is
reproduced throughout the displayed range, and both phase shifts are
accurately reconstructed. The largest differences occur near
$E\simeq5~\MeV$, where the observables vary rapidly and the inversion is
particularly sensitive to small numerical changes in the confined levels and
trap functions. The spherical-wall points lie somewhat closer to the
reference curves in this region, but both extractions remain in good overall
agreement with the continuum calculation.

Periodic-box results are not shown for the charged calculation. A consistent
treatment would require both a periodic formulation of the Coulomb interaction
and the corresponding Coulomb-modified finite-volume QC, neither of which is
included in the present benchmark.

%---------------------------
\subsection{$^3P_1$ results}
%---------------------------

\begin{figure*}[t]
\centering
\includegraphics[width=0.49\textwidth]{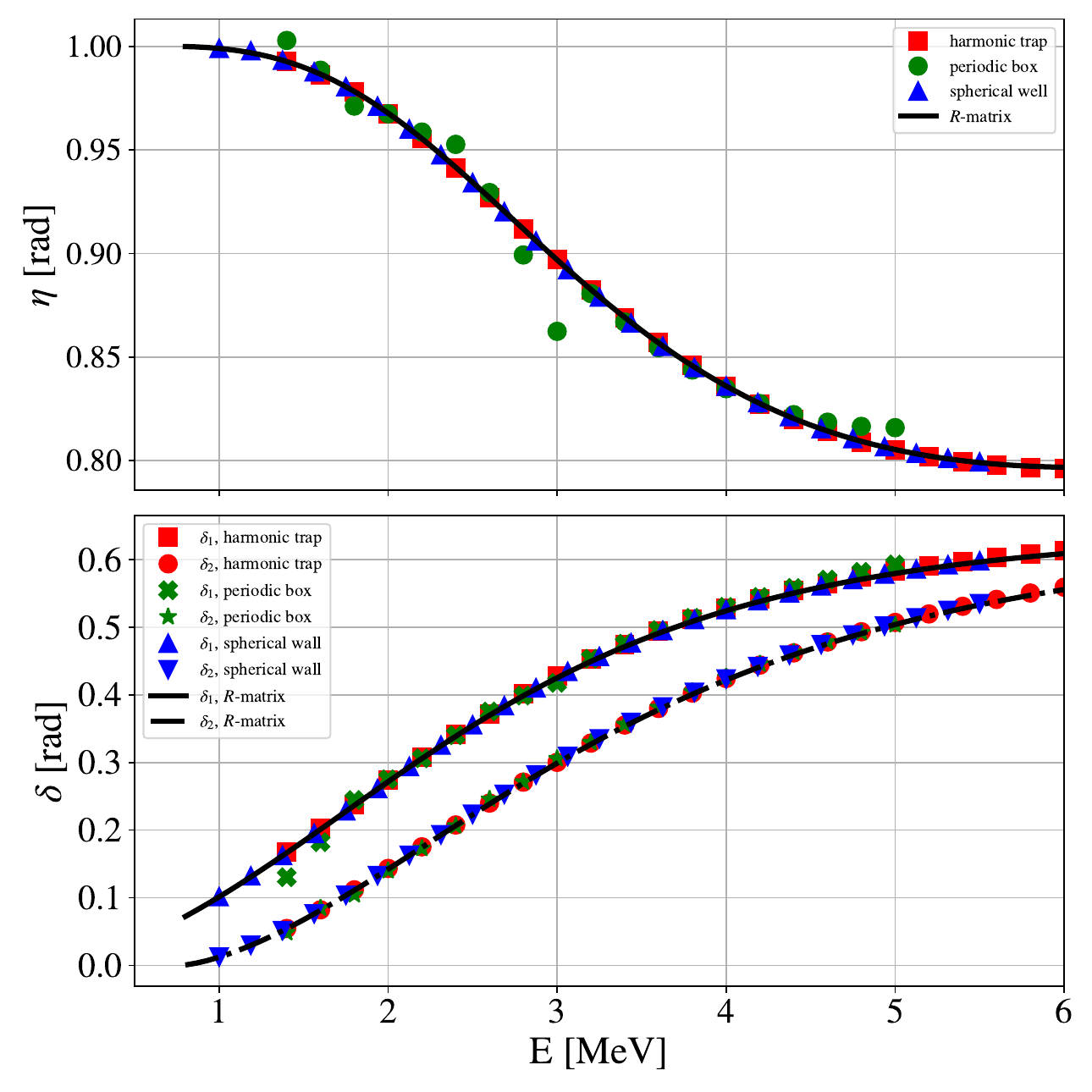}
\hfill
\includegraphics[width=0.49\textwidth]{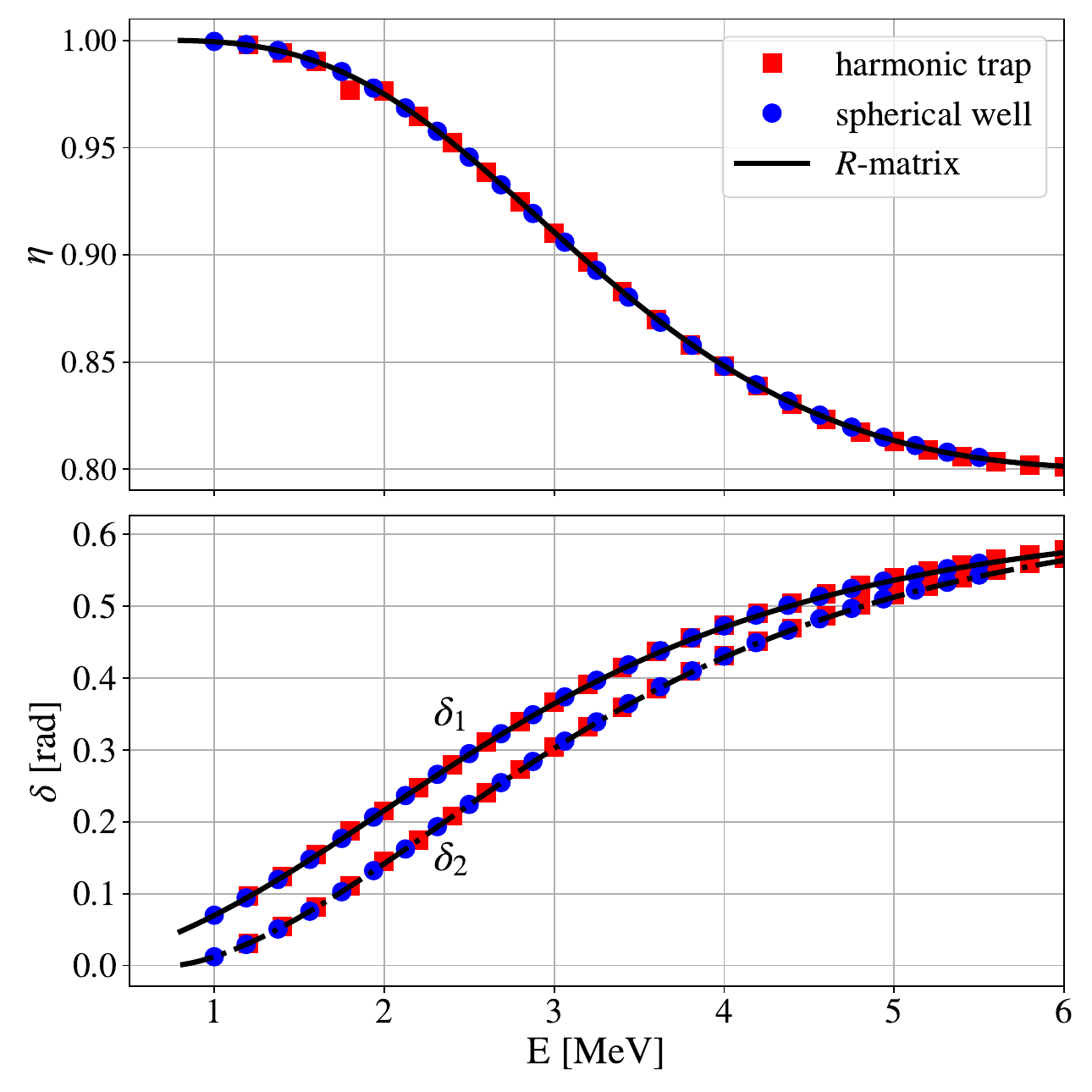}
\caption{Same as Fig.~\ref{fig:results_1s0}, but for the $^3P_1$ partial wave.}
\label{fig:results_3p1}
\end{figure*}

The $^3P_1$ results are shown in Fig.~\ref{fig:results_3p1}. In contrast to
the $^1S_0$ case, the inelasticity remains close to unity over the displayed
range, decreasing smoothly from values near one at low energy to approximately
$0.8$ at the upper end. Both phase shifts increase with energy, with
$\delta_1$ larger than $\delta_2$ throughout the interval.

Without Coulomb interactions, the HO, spherical-wall, and periodic-box
extractions closely reproduce the reference phase shifts and inelasticity.
The results from the three geometries are mutually consistent and follow the
energy dependence of the $R$-matrix calculation over the full interval. The
periodic-box inelasticity shows slightly greater scatter at a few intermediate
energies. These points lie near poles of the finite-volume trap function,
where small changes in the interpolated levels produce comparatively large
changes in $\mathcal{F}^{\rm box}$ and hence in the extracted observables. The
deviations remain small, and the periodic-box results are consistent with both
the other confinement schemes and the continuum reference. Minor differences
in individual phase-shift points also appear at the highest energies.

When Coulomb interactions are included, the HO and spherical-wall results
remain in excellent agreement with the $R$-matrix reference throughout the
displayed interval. Both phase shifts and the inelasticity are accurately
reproduced, demonstrating the consistency of the Coulomb-modified trap
functions for the charged-channel observables in the $^3P_1$ partial wave.

%-----------------------------------------------------------------
\subsection{Sensitivity to uncertainties in the confined spectrum}
\label{sec:sensitivity}
%-----------------------------------------------------------------

Uncertainties in the confined energies propagate through the interpolation to
common scattering energies, the evaluation of the trap functions, and the
nonlinear inversion of the QC. Spectral perturbations can therefore be
amplified at several stages, particularly near trap-function poles and along
poorly conditioned directions of the inverse problem. We quantify this
sensitivity with a Monte Carlo propagation study for the $^1S_0$ partial wave
without Coulomb interactions, for which all three geometries can be compared
on the same footing.

Each calculated confined energy is independently perturbed by sampling from a
Gaussian distribution with a relative standard deviation of either $1\%$ or
$0.1\%$.
For each realization, one perturbed version of the complete spectrum is used
at every target energy, thereby preserving correlations among the extracted
observables. The perturbed levels are represented by the same PCHIP
construction used in the central analysis. The crossings $\lambda_a$ at a
prescribed energy $E$ are relocated within neighborhoods of their unperturbed
values, and crossings too close to the edge of the computed $\lambda$ grid are
discarded. The trap functions are then reevaluated, and
$(\delta_1,\delta_2,\eta)$ are obtained from the same overdetermined fit as in
the central analysis. A short pilot ensemble is used to estimate the
covariance matrix of the QC residuals, whose inverse defines the weight matrix
$\bm W$ in Eq.~\eqref{eq:least_squares}.

Because the two-channel QC is multivalued, each fit is initialized from several
seeds based only on spectral information: the unperturbed-spectrum extraction
at the same energy; the accepted solution at the preceding energy of the same
noisy realization, when available; a physical low-energy prior with $\eta$
near unity; and a seed with the two phases interchanged. Every seed is allowed
to converge over the full physical interval $0\leq\eta\leq1$, without a search
box centered on the $R$-matrix solution. The converged minimum with the lowest
QC cost is retained. Missing crossings, nonfinite trap functions, and
nonconverged fits are recorded separately. The fraction of accepted
realizations remains close to unity throughout the displayed interval. The
branch-ambiguous fraction is only a few percent and increases appreciably only
near $4~\MeV$.

\begin{figure}[t]
    \centering
    \includegraphics[width=\columnwidth]{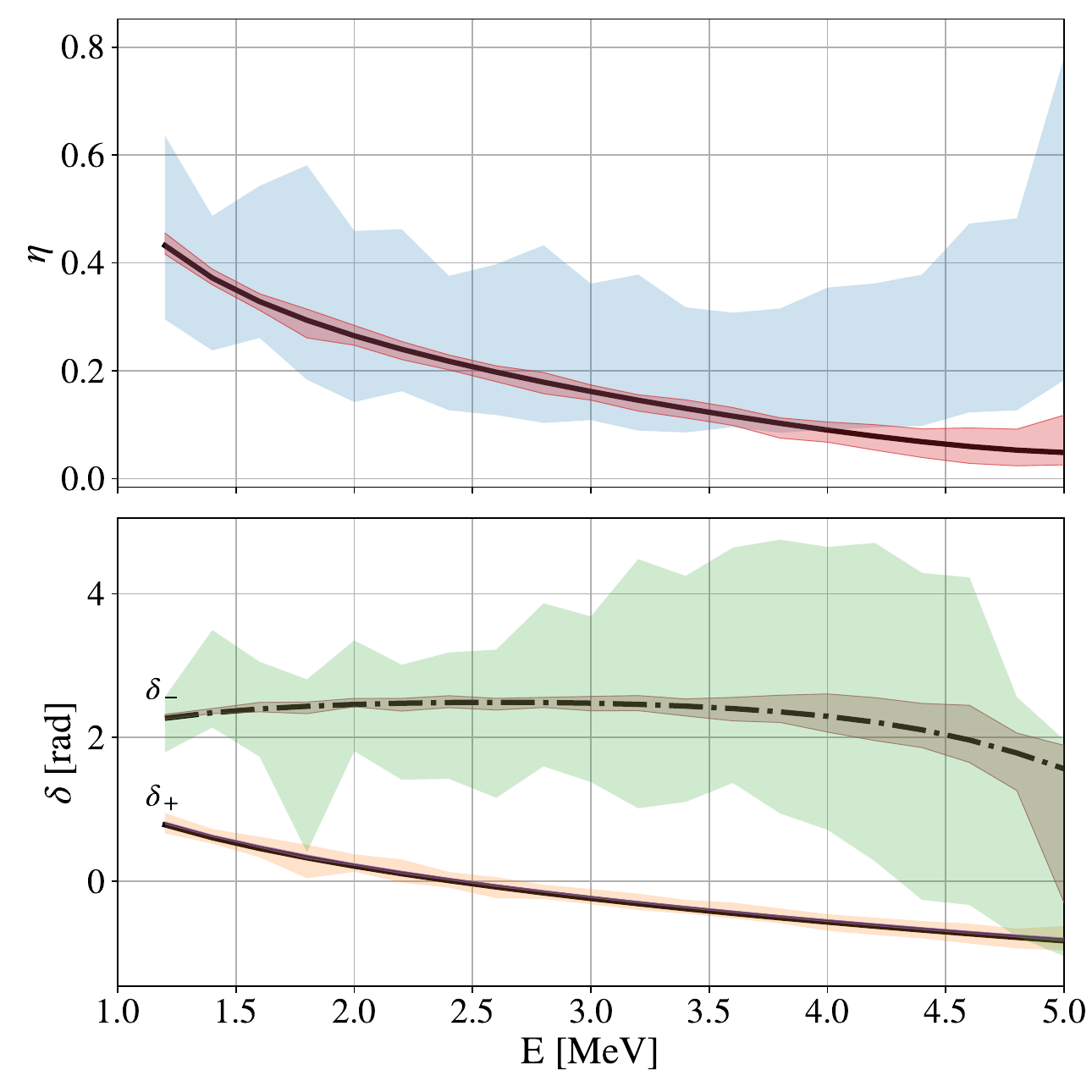}
    \caption{Propagation of spectral uncertainties to the extracted
    $^1S_0$ observables without Coulomb interactions for the HO trap. Relative
    Gaussian relative uncertainties of $1\%$ and $0.1\%$ are applied to the
    confined energies.
    The upper panel shows the inelasticity $\eta$, and the lower panel shows
    $\delta_{\pm}=\delta_1\pm\delta_2$. Solid curves are the $R$-matrix
    reference; shaded regions are the $16$--$84\%$ Monte Carlo intervals from
    $2\times10^5$ realizations. At each energy, the same number of constraints
    is used for all three geometries.}
    \label{fig:HO_sensitivity}
\end{figure}

\begin{figure}[t]
    \centering
    \includegraphics[width=\columnwidth]{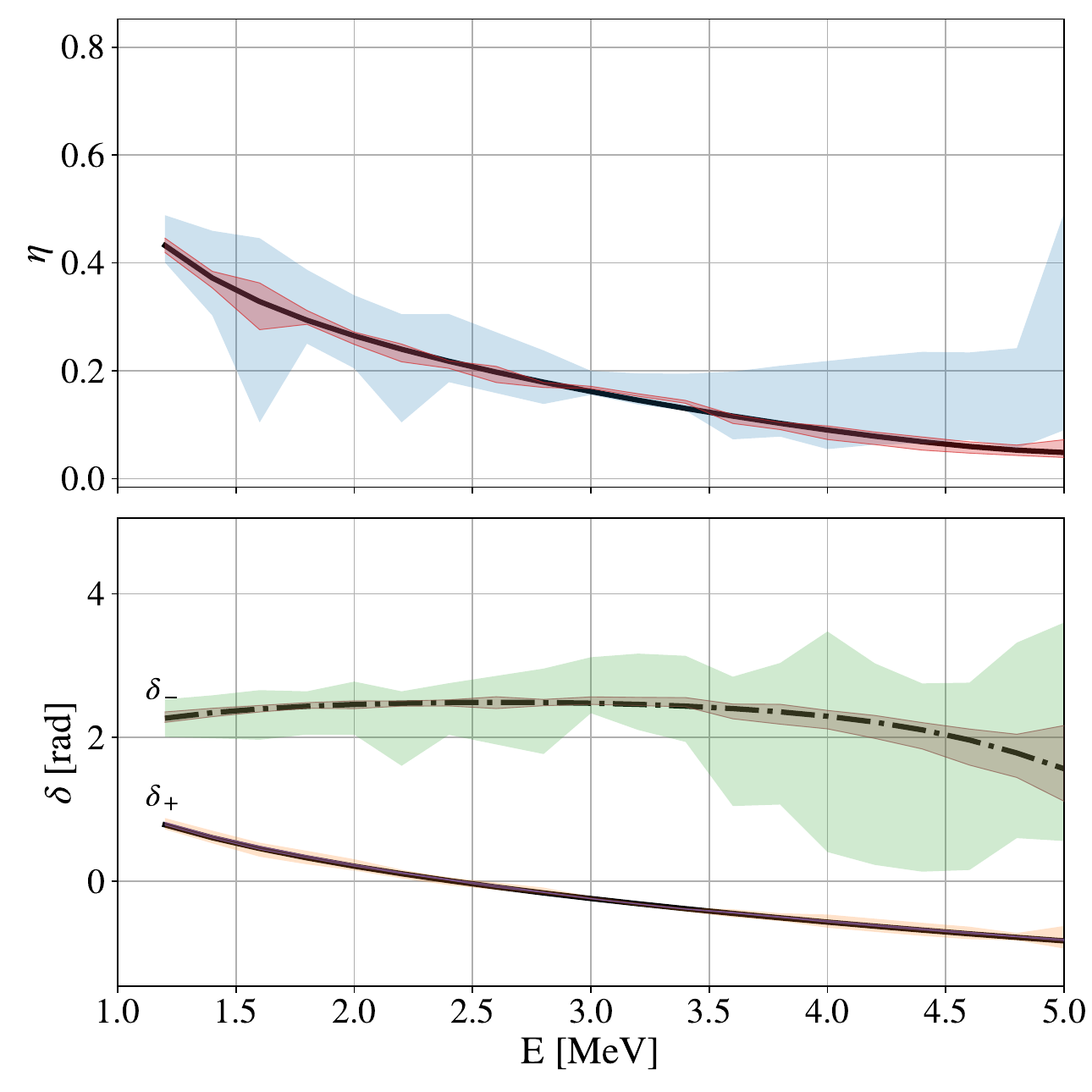}
    \caption{Same as Fig.~\ref{fig:HO_sensitivity}, but for the periodic cubic
    box.}
    \label{fig:BOX_sensitivity}
\end{figure}

\begin{figure}[t]
    \centering
    \includegraphics[width=\columnwidth]{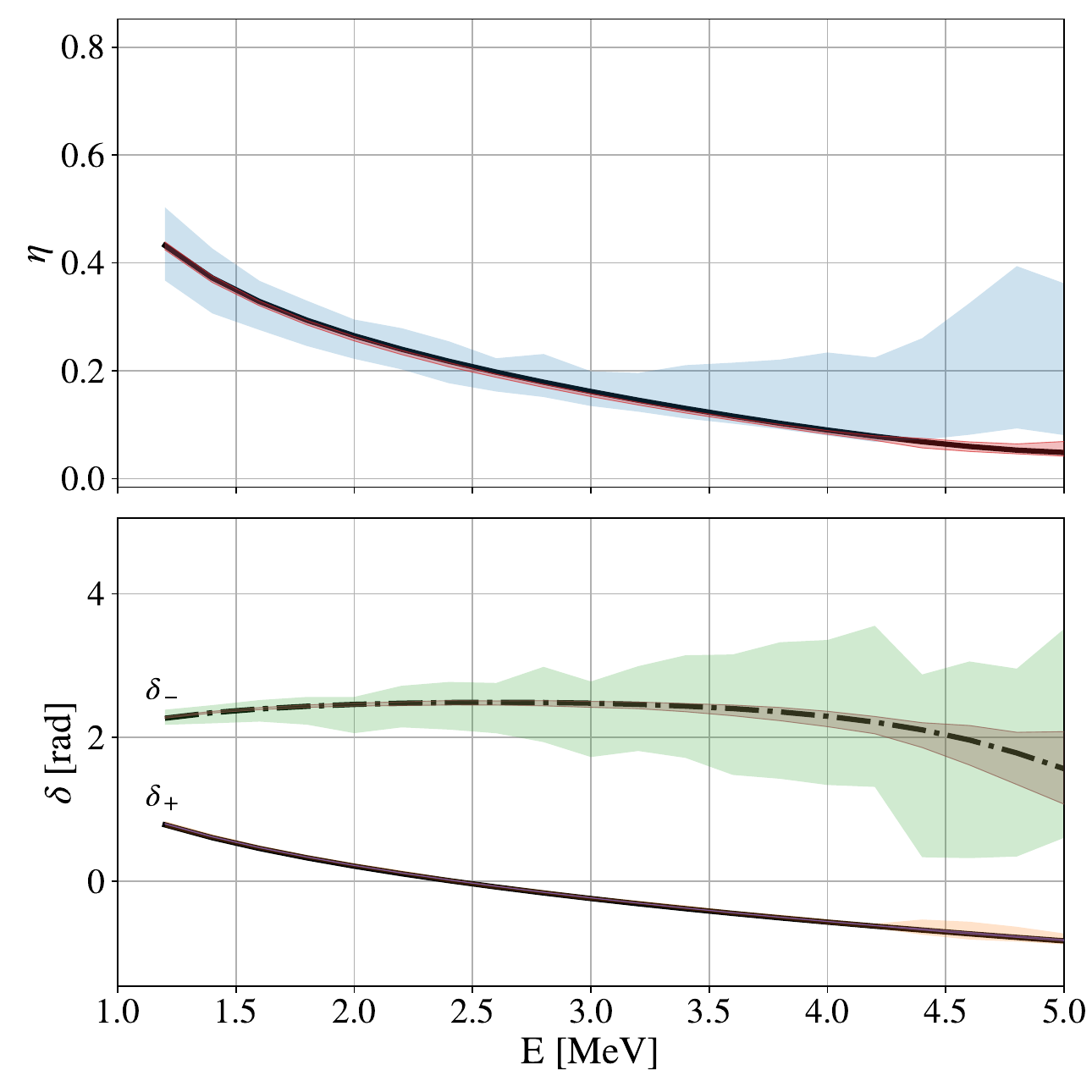}
    \caption{Same as Fig.~\ref{fig:HO_sensitivity}, but for the spherical hard
    wall.}
    \label{fig:SW_sensitivity}
\end{figure}

The number of usable crossings differs among the geometries. To compare them
at a common information content, we retain at each energy the same number of
constraints in all three schemes, equal to the smallest number of usable
crossings among them. Energies for which any geometry has fewer than three
crossings are omitted. Ensembles of $N_{\rm MC}=2\times10^5$ realizations are
then used to construct the bands shown in
Figs.~\ref{fig:HO_sensitivity}--\ref{fig:SW_sensitivity}. The solid curves
denote the $R$-matrix reference results, whereas the shaded regions indicate
the 16th--84th percentile intervals of the accepted samples.

We display the phase-shift combinations
\begin{equation}
    \delta_{\pm}(E)=\delta_1(E)\pm\delta_2(E).
    \label{eq:phase_shift_combinations}
\end{equation}

Figure~\ref{fig:HO_sensitivity} shows the HO results. With $0.1\%$ spectral
uncertainty, $\eta$ and $\delta_+$ follow the $R$-matrix reference through most
of the interval. The difference $\delta_-$ is the least stable combination:
its band remains useful to about $4~\MeV$ and then widens sharply toward
$5~\MeV$. At $1\%$ uncertainty, the $\eta$ and $\delta_-$ bands broaden at
lower energies. Near $3~\MeV$, the $1\%$ interval for $\eta$ is already several
times wider than the $0.1\%$ interval; by $4$--$5~\MeV$, the $1\%$ ensemble
shifts upward in $\eta$, while $\delta_-$ spans several radians. The sum
$\delta_+$ remains comparatively well determined at both noise levels.

The periodic-box results in Fig.~\ref{fig:BOX_sensitivity} exhibit the same
hierarchy: $\delta_+$ is robust, whereas $\eta$ and $\delta_-$ are sensitive.
The $1\%$ bands nevertheless remain narrower than their HO counterparts up to
about $4~\MeV$. An upward shift in $\eta$ and rapid broadening of $\delta_-$
appear mainly near $5~\MeV$, where the finite-volume trap function changes
rapidly in the vicinity of its poles. At $0.1\%$ uncertainty, both $\eta$ and
$\delta_+$ track the reference closely, although $\delta_-$ still broadens at
the upper end of the interval. A few percent of the $1\%$ samples are
branch-ambiguous near $4~\MeV$.

With the number of constraints matched, the spherical-wall extraction in
Fig.~\ref{fig:SW_sensitivity} is similar to the periodic-box result and more
stable than the HO extraction under $1\%$ noise. The wall function
$\mathcal{F}^{\rm SW}=n_\ell/j_\ell$ has poles at the zeros of $j_\ell$, but
they are more widely spaced than the noninteracting HO poles. The $0.1\%$
bands for $\eta$ and $\delta_+$ remain narrow throughout the displayed range,
and $\delta_+$ retains useful precision even at $1\%$. As in the other
geometries, $\delta_-$, followed by $\eta$ at the larger noise level, degrades
first as the energy increases and the inelasticity decreases.

Across all three geometries, reducing the relative spectral uncertainty from
$1\%$ to $0.1\%$ substantially improves both the accuracy and precision of the
extraction. The phase-shift sum $\delta_+$ is well conditioned throughout the
displayed interval, whereas $\eta$ and especially $\delta_-$ require
sub-percent spectral precision above approximately $4~\MeV$ in this benchmark.
This hierarchy is a property of the two-channel inverse problem: as $\eta$
becomes small, the QC continues to constrain $\delta_+$ tightly but becomes
nearly degenerate in the $(\eta,\delta_-)$ plane. Once the number of
constraints is matched, the differences among the geometries are milder than
a comparison of the raw extractions would suggest. The HO $1\%$ bands still
broaden at lower energies, consistent with the denser noninteracting poles,
but this comparison applies to the present grids and a common relative energy
uncertainty, not to separately optimized many-body implementations. The loss
of precision in $\delta_-$ at high energy is common to all three schemes.

%======================
\section{Conclusions}
\label{sec:conclusion}
%======================

We have investigated the extraction of coupled-channel scattering observables
from spectra generated by an HO trap, a spherical hard wall, and a periodic
cubic box. The \tHp--\tHen\ cluster model of \HeFour\ provides a controlled
benchmark because it combines a nearby second-channel threshold with a
long-range Coulomb interaction in only one channel.

The three confinement schemes admit a common description in which the geometry
enters through channel-dependent trap functions. Above the \tHen\ threshold,
each confined level provides one constraint on two phase shifts and an
inelasticity. Several independent crossings at the same scattering energy are
therefore required to determine the three observables, and an overdetermined
fit improves the stability of the inversion.

Without Coulomb interactions, the HO, spherical-wall, and periodic-box
extractions are mutually consistent and agree closely with the continuum
$R$-matrix reference for both partial waves. The largest deviations occur near
trap-function poles, in regions where the observables vary rapidly, or when
the inverse problem becomes poorly conditioned. With Coulomb interactions,
the HO and spherical-wall methods continue to reproduce the continuum
reference over the energy range considered. The spherical wall permits direct
Coulomb matching, whereas the HO geometry is naturally compatible with
oscillator-basis many-body calculations. Its inverse extraction nevertheless
requires particular care near the noninteracting poles. The periodic box is
closely connected to lattice and finite-volume calculations; in the present
analysis, however, partial waves beyond the leading contribution in each
cubic irrep are neglected.

The uncertainty analysis reveals a clear hierarchy among the extracted
quantities. The phase-shift sum $\delta_+$ remains comparatively well
conditioned, whereas the inelasticity and especially the phase-shift difference
$\delta_-$ are sensitive to the precision of the confined spectrum. Reducing
the relative spectral uncertainty from $1\%$ to $0.1\%$ markedly improves the
higher-energy extraction of $\eta$ and $\delta_-$. At matched information
content, the HO bands under $1\%$ noise broaden earlier than those of the box
and spherical wall, consistent with the denser noninteracting poles of the
oscillator trap. The remaining loss of precision in $\eta$ and $\delta_-$ at
high energy is common to all three geometries and follows from the
near-degeneracy of the coupled-channel QC when the inelasticity is small.

No single confinement geometry is optimal for every application. The
appropriate choice depends on the basis or spatial representation used to
calculate the spectrum, the treatment of long-range Coulomb interactions, and
the relevance of higher partial waves. The present comparison establishes a
common benchmark for spectrum-based coupled-channel extractions and provides a
foundation for extending these methods to more realistic few-body and
\textit{ab initio} reaction calculations.

\begin{acknowledgments}
We thank Nir Barnea for useful discussions and comments.  
This work was supported by the Israel Science Foundation (ISF), Grant
No.~2441/24.
\end{acknowledgments}

\end{document}